\documentclass[twocolumn]{aastex631}

\usepackage{graphicx}
\usepackage{float}
\usepackage{mathrsfs}	
\usepackage{amsmath}
\usepackage{color}
\usepackage{multirow}
\usepackage{subfigure}
\usepackage{longtable}
\usepackage{blindtext}
\usepackage{amssymb}
\def\gsim{~\rlap{$>$}{\lower 1.0ex\hbox{$\sim$}}}
\def\lsim{\mathrel{\rlap{\lower2.5pt\hbox{\hskip0.5pt$\sim$}}
    \raise0.8pt\hbox{$<$}}} 

\newcommand{\rui}[1]{{\textcolor{black}{#1}}}
\newcommand{\ruili}[1]{{\textcolor{black}{#1}}}
\shorttitle{AASTeX v6.3.1 Sample article}
\shortauthors{Li et al.}
\graphicspath{{./}{figures/}}
\begin{document}

%\title{A bright Einstein ring lensed by massive isolated galaxy at redshift $z\sim1$}
\title{Multi-band analysis of strong gravitationally lensed post-blue nugget candidates from the Kilo-Degree Survey}\shorttitle{Strong lensed post-blue nugget from KiDS}

\author[0000-0002-3490-4089]{Rui Li}
\affiliation{Institude for Astrophysics, School of Physics, Zhengzhou University, Zhengzhou, 450001, China.
{\it \rm liruiww@gmail.com}}
\affiliation{National Astronomical Observatories, Chinese Academy of Sciences, 20A Datun Road, Chaoyang District, Beijing 100012, China}
\author{Nicola R. Napolitano}
\affiliation{Department of Physics ``E. Pancini'', University Federico II, Via Cinthia 6, 80126-I, Naples, Italy
{\it \rm nicolarosario.napolitano@unina.it}}
\affiliation{School of Physics and Astronomy, Sun Yat-sen University, Zhuhai Campus, 2 Daxue Road, Xiangzhou District, Zhuhai, P. R. China.} 
\affiliation{CSST Science Center for Guangdong-Hong Kong-Macau Great Bay Area, Zhuhai, 519082, P. R. China}
\author{ Linghua, Xie}
\affiliation{School of Physics and Astronomy, Sun Yat-sen University, Zhuhai Campus, 2 Daxue Road, Xiangzhou District, Zhuhai, P. R. China}
\author{ Ran Li}
\affiliation{National Astronomical Observatories, Chinese Academy of Sciences, 20A Datun Road, Chaoyang District, Beijing 100012, China}
\affiliation{School of Astronomy and Space Science, University of Chinese Academy of Sciences, Beijing 100049, China}
\author{Xiaotong Guo}
\affiliation{Institute of Astronomy and Astrophysics, Anqing Normal University, Anqing, Anhui 246133, China}
%\author{ Ran Li*}
\author{Alexey Sergeyev}
\affiliation{Universit\'{e} C\^ote d’Azur, Observatoire de la C\^ote d’Azur, CNRS, Laboratoire Lagrange, France}
\affiliation{V. N. Karazin Kharkiv National University, Kharkiv, 61022, Ukraine}
\author{ Crescenzo Tortora}
\affiliation{INAF -- Osservatorio Astronomico di Capodimonte, Salita Moiariello 16, 80131 - Napoli, Italy}
\author{ Chiara Spiniello}
\affiliation{Department of Physics, University of Oxford, Denys Wilkinson Building, Keble Road, Oxford OX1 3RH, UK}
\affiliation{INAF -- Osservatorio Astronomico di Capodimonte, Salita Moiariello 16, 80131 - Napoli, Italy}
\author{Alessandro Sonnenfeld}
\author{L\'{e}on V. E. Koopmans}
\affiliation{Kapteyn Astronomical Institute, University of Groningen, P.O.Box 800, 9700AV Groningen, the Netherlands}
\author{Diana Scognamiglio}
\affiliation{Jet Propulsion Laboratory, California Institute of Technology,  4800, Oak Grove Drive - Pasadena, CA 91109, USA}
%\author{Ran Li}
%\affiliation{School of Astronomy and Space Science, University of Chinese Academy of Sciences, Beijing 100049, China}
%\affiliation{National Astronomical Observatories, Chinese Academy of Sciences, 20A Datun Road, Chaoyang District, Beijing 100012, China}

%\author{ Linghua Xie}
%\affiliation{School of Physics and Astronomy, Sun Yat-sen University, Zhuhai Campus, 2 Daxue Road, Xiangzhou District, Zhuhai, P. R. China}



\begin{abstract}During the early stages of galaxy evolution, a significant fraction of galaxies undergo a transitional phase between the ``blue nugget" systems, which arise from the compaction of large, active star-forming disks, and the ``red nuggets", which are red and passive compact galaxies. These objects are typically only observable with space telescopes, and detailed studies of their size, mass, and stellar population parameters have been conducted on relatively small samples. Strong gravitational lensing can offer a new opportunity to study them in detail, even with ground-based observations. In this study, we present the first 6 \textit{bona fide} sample of strongly lensed post-blue nugget (pBN) galaxies, which were discovered in the Kilo Degree Survey. By using the lensing-magnified luminosity from optical and near-infrared bands, we have derived robust structural and stellar population properties of the multiple images of the background sources. \rui{The pBN galaxies have very small sizes of $\rm R_{eff}<1.3$ kpc, high mass density inside 1 kpc of $ \rm \log (\Sigma_1/M_{\odot} \mathrm{kpc}^{-2})>9.3$, and low specific star formation rates of $\log (\mathrm{sSFR/Gyr})\lesssim0$,
%(except only J023929$-$321129)
}
%($\log \mathrm{sSFR/Gyrs}\lesssim0.5$),
%($\log \mathrm{sSFR/Gyrs}\lesssim0$),
%which places them between the blue and red nugget phases. 
The size-mass and $\Sigma_1$-mass relations of this sample are consistent with those of the red nuggets, while their sSFR is close to the lower end of compact star-forming blue nugget systems at the same redshift, suggesting a clear evolutionary link between them.

\end{abstract}

\keywords{Strong lensing --- galaxies --- Einstein ring}


\section{Introduction} \label{sec:intro}
According to the hierarchical galaxy formation scenario (e.g., \citealt{Renzini2006ARA&A..44..141R, Lapi2018ApJ...857...22L}), large primordial disk galaxies undergo a dissipative process in which gas streams from the cosmic web fall into the disks, concentrate in the central region, and lead to a high rate of star formation (\citealt{Dekel2009ApJ...703..785D}). This process results in the formation of extremely compact and star-forming galaxies (SFGs), known as ``blue nuggets" (\citealt{Barro2013ApJ...765..104B, Barro014ApJ...791...52B}, \citealt{Zolotov2015MNRAS.450.2327Z}, \citealt{Tacchella2016MNRAS.458..242T}). They are characterized by a very small effective radius, $\rm R_{eff}<1 \ kpc$ (e.g., \citealt{Williams2014ApJ...780....1W}; \citealt{Zolotov2015MNRAS.450.2327Z}), high surface density inside the central 1 kpc, $\rm log (\Sigma_1/\ M_{\odot} \mathrm{kpc}^{-2})>9.0$ (e.g. \citealt{Barro2013ApJ...765..104B}; \citealt{Dekel2014MNRAS.438.1870D}), and a high specific star-forming rate, $\rm log(sSFR/Gyr)>0$ (e.g., \citealt{Zolotov2015MNRAS.450.2327Z}; \citealt{Huertas-Company2018ApJ...858..114H}). The lifetime of these blue nuggets can be quite short ($\rm <1\ Gyr$), as they are expected to quickly switch off their star formation
%($\rm log\ sSFR/Gyr \lesssim 0 $)
and turn into post-blue nuggets (pBNs) of similar structural properties (\citealt{Dekel2014MNRAS.438.1870D, Zolotov2015MNRAS.450.2327Z}) but lower sSFR of $\rm log(sSFR/Gyr) \lesssim 0 $. The actual mechanisms behind these ``quick'' transformations are yet to be clarified. They are thought to consist of the combined effect of feedback from star winds induced by strong starbursts and the presence of Active Galactic Nuclei (AGN, e.g., \citealt{Dekel1986ApJ...303...39D, Murray2005ApJ...618..569M, Zolotov2015MNRAS.450.2327Z, Cattaneo2009Natur.460..213C}). These mechanisms should be effective over a short timescale ($\sim$ 1 Gyr or less) and at higher redshifts ($z>2$, \citealt{Dekel2014MNRAS.438.1870D, Costantin2021ApJ...913..125C}). After complete quenching, the pBNs turn into compact red nuggets, which finally grow in mass and size through dry merging, forming present-day elliptical galaxies  (e.g., \citealt{Newman2012ApJ...746..162N, Oser2012ApJ...744...63O, Nipoti2012MNRAS.422.1714N, Posti2014MNRAS.440..610P, Spiniello2021A&A...646A..28S}).

If the red nuggets indeed evolved from blue nuggets, there must be a continuum between some basic properties (e.g. stellar mass--$\rm M_*$, age, specific star-forming rate--$\rm sSFRs$, effective radius--$\rm R_{eff}$ and surface density inside 1 kpc--$\rm \Sigma_1$, see e.g. \citealt{Tacchella2016MNRAS.458..242T}) of blue nuggets, pBNs and red nuggets (see e.g. \citealt{Huertas-Company2018ApJ...858..114H}). Hence studying the relationships between these properties (e.g. $\rm R_{eff}-M_*$, $\rm sSFR-M_*$, and $\rm \Sigma_1-M_*$)
%, e.g. \citealt{Barro014ApJ...791...52B, Barro2017ApJ...840...47B}) 
%(e.g. size-mass, SFR-$\Sigma_1$, and size-$\Sigma_1$, e.g. \citealt{Barro014ApJ...791...52B, Barro2017ApJ...840...47B}) 
can help us better understand their evolutionary path and determine their roles in the galaxy evolution scenario. This evolutionary track has been investigated in simulations by \cite{Zolotov2015MNRAS.450.2327Z}, showing that the majority of compaction of blue nuggets begins when $\rm log (M_*/M_{\odot})<9.5$ and $\rm R_{eff}>1\ kpc$, followed by quenching $\rm log (M_*/M_{\odot})\sim 10$ and $\rm R_{eff}<1\ kpc$. This compaction phase occurs up to $z\sim2$. Furthermore, galaxies with larger masses experience quenching at a higher value of $\Sigma_1$, whereas lower mass galaxies quench at lower values of $\Sigma_1$. As blue nuggets transit into red nuggets, their size begins to increase linearly with mass along the passive galaxy size-mass relation.
%\red{the mass and size begin to increase through mergers and accretion.}
%the size-mass relation begins to increase linearly. 
On the observational side, the size-mass relation of red nuggets has been investigated through direct observations (e.g. \citealt{Damjanov2009ApJ...695..101D}) or the strong lensing magnification effect (\citealt{Oldham2017MNRAS.465.3185O}). Although a few blue nuggets have also been found and confirmed (\citealt{Marques-Chaves2022MNRAS.517.2972M}), their size-mass relation remains poorly constrained (see e.g. \citealt{Williams2014ApJ...780....1W}), mainly
because the sizes of this kind of galaxies are difficult to measure.

Strong lensing (SL), can provide powerful observational support to systematically study these objects. The warped geometry of the space and time generated by the gravitational field of massive objects (lenses) in the universe, produces a magnification of the luminosity of the background ``lensed'' galaxies. This is particularly efficient for compact sources, which can be studied in much higher detail (\citealt{Oldham2017MNRAS.465.3185O}), even with ground-based observations (\citealt{Napolitano2020ApJ...904L..31N}).
%which refers to the phenomenon where light emitted from a background source undergoes significant bending when passing through a massive foreground deflector due to the gravitational potential is a powerful tool for investigating faint, high-redshift galaxies. This is because the SL can magnify the flux of background sources, making it easier to measure the flux of faint objects and even reconstruct their structures. The use of SL enables the analysis of nugget galaxies to reach unprecedented levels of detail, utilizing data from both space and ground-based observations.
%\rui{Strong lensing (SL), which refers to the phenomenon where light emitted from a background source undergoes significant bending when passing through a massive foreground deflector due to the gravitational potential, is a natural telescope to help study the high redshift fainter galaxies. This is because SL magnifies the flux of background sources, making it easier to measure the flux of faint objects and even reconstruct their structures. 
%This is because SL has a magnifying effect on background sources, which makes it easier to measure the flux of faint objects, and even reconstruct their own structures. 
%With SL effect, the analysis of nugget galaxies can be pushed to reach many finest details using data from space and even from the ground.}
%Strong lensing(SL) can push this kind of analysis to reach many finest details using data from space and even from the ground.
For example, \citet{Toft2017Natur.546..510T} analyzed a strongly lensed red nugget and suggested that the stars in this galaxy formed in a disk rather than a merger-driven nuclear starburst. \rui{Two Einstein crosses from pBNs have been discovered and confirmed with VLT/MUSE observations by \citet[][hereafter N+20]{Napolitano2020ApJ...904L..31N} in the footprint of the Kilo Degree Survey (KiDS, \citealt{deJong2013ExA....35...25D}). This finding has shown the path for systematic studies of pBNs via quadruple lensing events in future ground and space large sky surveys, where thousands of similar systems have been predicted to be found (see the discussion in N+20).}
\rui{Despite in N+20 we have estimated the expected number of the lensed pBNs in KiDS \ruili{(namely, half a dozen for sources at $z<4$)}, there we did not concentrate on all  
%primarily concentrated on analyzing lensed pBNs that had been spectroscopically confirmed, thereby not including all 
potential candidates in the surveyed area, but only on the two pBNs that had been spectroscopically confirmed. In this work, we instead add 
%the KiDS survey's pBN sample with the addition of four 
all previously unexplored candidates, despite the absence of spectroscopic data for them. The purpose of this paper is to fully introduce 
%we are able to employ 
a comprehensive optical and near-infrared ray tracing analysis that, without spectroscopy, can allow us to derive the photometric redshifts (photo-$z$s) for both the foreground lensing galaxy and the background source, thus photometrically confirming the lensing nature of the systems. 
%This sample is consistent with the number of events predicted in the full KiDS area from N+20. Hence, the primary objective of 
Here we also discuss that, besides confirmation, the method can allow us to characterize the properties of the bakground sources
%. This will further allow us to compare    
and derive their scaling relations that we can use to compare with the one of the red nuggets and discuss an evolutionary link with them. 
%with samples of blue and red nuggets from previous literature and find proof of their evolutionary link.
}

The paper is organized as follows. Section 2 describes the data used in this study. Section 3 outlines the main methods used for lens modelling, photo-z estimation, and determination of stellar population parameters. Section 4 discusses the multi-band ray-tracing results and the scaling relations of the six lensed pBNs. Finally, Section 5 concludes our study. For all calculations, we assume a $\Lambda$CDM cosmology with ($\Omega_M$, $\Omega_\Lambda$, $h$)=(0.3, 0.7, 0.7).

\begin{figure}
    \centering
    \includegraphics[width=8.5cm]{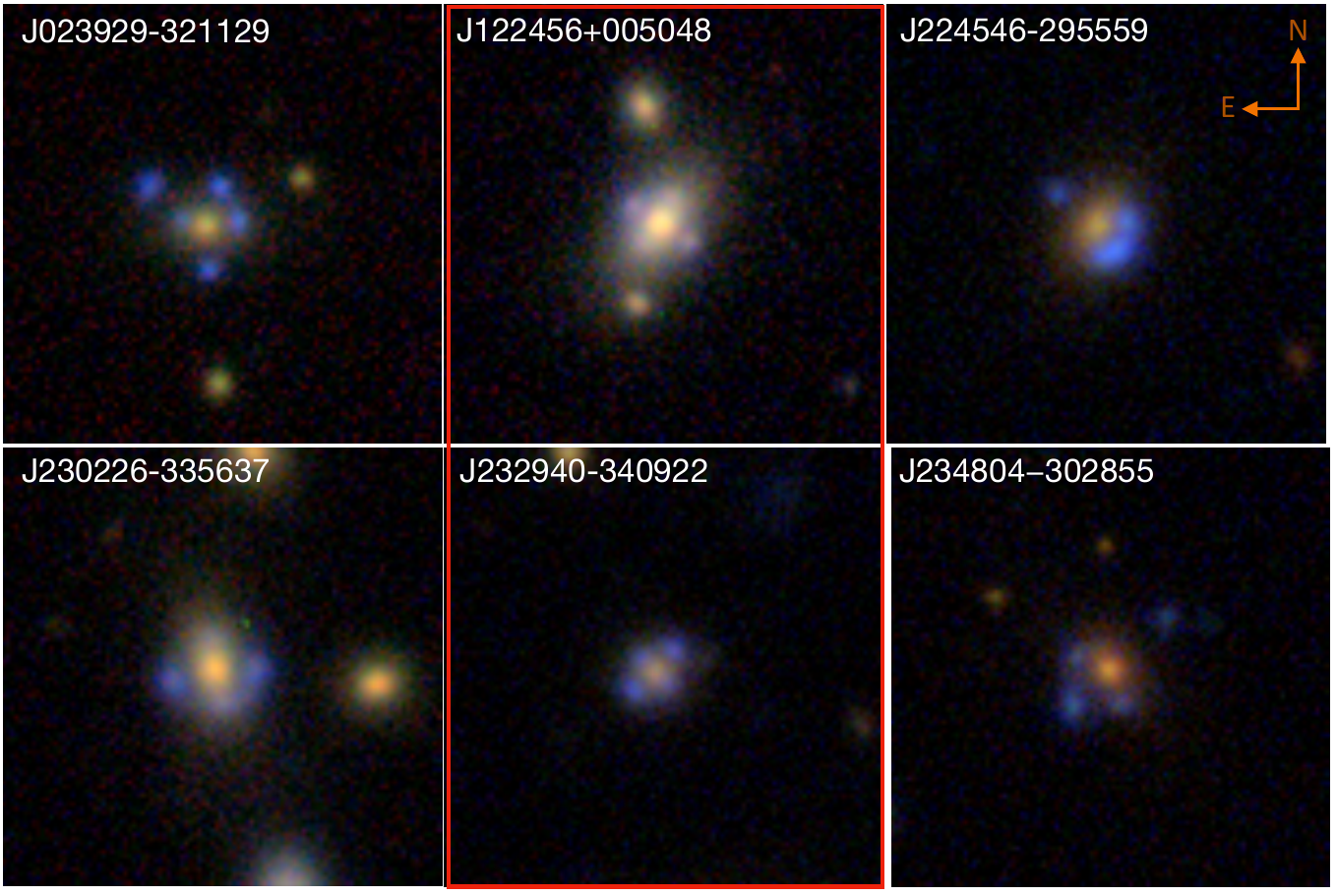}\vspace{3pt}
    \caption{Colored stamps ($20'' \times 20''$) of the six lensed blue nuggets, obtained by combining {\it g, r, and i-}band images from KiDS. Each system exhibits three or four point-like images, indicating that the background sources may be very small in size. {The two systems highlighted in red have been confirmed with MUSE/VLT spectroscopy in N+20.} }
    \label{fig:color_images}
\end{figure}

\section{Data}
\label{target_selection} 

\rui{The data used in this analysis consist of optical images from the KiDS and near-infrared (NIR) images from the VISTA Kilo-degree Infrared Galaxy (VIKING) surveys. KiDS covers a total area of 1350 deg$^2$ in four optical filters ($u, g, r, i$). The seeing, measured by the full width at half maximum (FWHM), ranges from approximately $0.7''$ to $1.1''$ across all the bands, with the r-band image being the sharpest ones (median $\rm FWHM \sim 0.7''$). The mean limiting AB magnitude ($5\sigma$ level within a $2''$ aperture) are 24.23 $\pm$ 0.12, 25.02 $\pm$ 0.13, 25.12 $\pm$ 0.14, and 23.68 $\pm$ 0.27, in the  $u, g, r, i$ bands respectively (see \citealt{Kuijken2019A&A...625A...2K} for more details).
The VIKING survey (see \citealt{Edge2013Msngr_VIKING}) covers the same area as KiDS but through five different NIR bands ($Z, Y, J, H, Ks$). The median seeing for these images is around $0.9''$ and the AB magnitude depths are 23.1, 22.3, 22.1, 21.5, and 21.2, in all filters, respectively (see \citealt{Sutherland2015A&A...575A..25S_VISTA} for more details).}

%Additionally, we perform a stellar population analysis of the sources by fitting the spectral energy distribution (SED) to determine the specific star-formation rate of the background galaxies and assess the pBN nature of the sources. The final sample of the {\it bona fide} pBNs is shown in Fig. \ref{fig:color_images}, which includes the two systems previously analyzed in N+20 (KiDSJ122456$+$005048 and KiDSJ232940$-$340922) and four new systems (KiDSJ023929$-$321129, KiDSJ224546$-$295559, KiDSJ230226$-$335637, and KiDSJ234804-302855).

\rui{In the KiDS survey, we have identified 268 high-quality SL candidates using mainly a Deep Learning approach (\citealt{Petrillo2019MNRAS.484.3879P, Li2020ApJ...899...30L, Li2021ApJ...923...16L}). To assemble a collection of lensed pBNs, we conducted a visual examination on the $gri$ composited stamps of these candidates. This process resulted in 14 SL candidates that potentially exhibit pBN sources. Each of them shows blue lensed images, with at least one image appearing point-like, indicating that the background sources are compact, similar to the lensed pBNs in N+20.}
%In KiDS, we have found 268 high-quality strong lens candidates with Machine Learning method (\citealt{Petrillo2019MNRAS.484.3879P, Li2020ApJ...899...30L, Li2021ApJ...923...16L}). To collect the lensed pBNs, we have visually inspected the $gri$ composited stamps of these lens candidates and found 14 of them potentially to be strong lenses with pBN sources. Each of these 14 candidates has blue lensed images, and at least one of the images is point-like, suggesting compact sizes of the background sources, similar to the lensed pBNs in N+20. 
Among these 14 candidates, two systems have more than one foreground lenses, which would make the lens modelling very uncertain. 
%For 6 of the remaining 12 candidates, we are hard to inspect their lensed images in at least 3 NIR bands, even after subtracting the foreground light before preforming the strong lens modelling.
\ruili{For 6 of the remaining 12 candidates, we encountered substantial difficulties in discerning the lensed images across at least three NIR bands, %This challenge was not mitigated even 
even after subtracting the foreground light before performing the strong lens modeling.}
%Furthermore, 6 of the remaining 12 candidates have an insufficient signal-to-noise ratio in at least three NIR bands.
%, which we need for our analysis. 
%This is partially because of the shallower depth of the VIKING data, but also for the combination of redshift and SED distribution of the background sources, which tend to be fainter in NIR if they are younger rather than older systems, at high-$z$ (see also in \citealt{Williams2014ApJ...780....1W}, Fig. 17).
\ruili{This is due partially because of the shallower depth of the VIKING data, but also of the properties of the SED distribution of high-z background sources, very likely young systems. 
%The brightness will tend to be fainter in NIR if the galaxies are 
Indeed, for younger 
%rather than older 
systems at high-$z$ we might expect a faint NIR (see also in \citealt{Williams2014ApJ...780....1W}, Fig. 17). In this respect, the visibility in NIR tend to naturally select the older systems, that are more compatible wth a pBN phase. In fact,}
the limiting magnitudes of different bands in the VIKING survey are $\rm mag_{AB}\lesssim23$ %Therefore, the 
and SFGs with rest-frame optical band magnitudes around 23--24 AB will be hard to detect, even if they are magnified by lensing. {\it This is an important selection effect to keep in mind in the following analysis.} Finally, we are left with 6 SL systems with optical+NIR imaging data good enough for the analysis we aim to perform in this work. 
\rui{In Fig. \ref{fig:color_images} we show the $gri$ color combined images of the final 6 SL candidates (KiDSJ023929$-$321129, KiDSJ122456$+$005048, KiDSJ224546$-$295559, KiDSJ230226$-$335637, KiDSJ232940$-$340922, and KiDSJ234804-302855) used for this analysis.} KiDSJ122456+005048 and KiDSJ232940-340922 are the two lensed pBNs observed with MUSE spectroscopy from N+20. We keep these ``confirmed'' cases in the sample to check if, with the pure photometric approach proposed here, we obtain consistent results in terms of the redshifts of the lenses and the sources. We also note that KiDSJ023929+321129 was studied by \citealt{Sergeyev2018RNAAS...2..189S} as a gravitational quadruple lens candidate, without giving the redshift of the sources.

\rui{The magnification effect of this SL sample, combined with the 9-band optical images from KiDS and VIKING survey, enable the accurate reconstruction of the light distribution of the background sources and the subsequent determination of their photo-$z$s. These data are the foundation of a detailed stellar population analysis via the spectral energy distribution (SED) fitting, that will finally provide us an estimate of the specific star-formation rate (sSFR) of the background galaxies and assess the pBN nature of the sources.}

\begin{table*}
\begin{center}
\scriptsize
\label{tab:source_mod}
\caption{\label{tab:source_mod} Source properties from the multi-band ray tracing ``flexible'' model}
\begin{tabular}{l c c c c c c c c c c c c c c c c}
\hline \hline
ID & $R_{\rm eff}$ & $b/a$ & $pa$ & $n_r$ & u & g & r & i & Z & Y & J & H & Ks \\
&(arcsec)  && (deg) & &(mag)&(mag)&(mag)&(mag)&(mag)&(mag)&(mag)&(mag)&(mag)\\
\hline
J023929$-$321129 & 0.063$\pm$0.002   & 0.40$\pm$0.02  & 89 & 2.7$\pm$0.1  & 23.70& 23.29& 23.14& 23.21&23.08& 23.44&22.48& 22.49& 22.11 \\
J122456$+$005048* & 0.057$\pm$0.002  & 0.53$\pm$0.04  & 163   &0.8$\pm$0.1  &-&24.05& 23.67& 23.14& 22.52& 22.24& 21.91& 21.88& 21.67\\
J224546$-$295559&  0.054$\pm$0.006   & 0.46$\pm$0.05  & 16    &3.4$\pm$0.5  & 24.02& 23.90&23.73& 23.72& 23.20&22.79&22.62&-&-\\
J230226$-$335637  & 0.154$\pm$0.008   & 0.52$\pm$0.05  & 61   &1.8$\pm$0.3  &23.38& 22.89& 22.96& 22.74& 22.94& 22.62& 22.14& 21.89& 22.56\\
J232940$-$340922* & 0.071$\pm$0.011  &  0.26$\pm$0.08  & 161  & 1.1 $\pm$0.2  &-&23.68& 23.52& 23.53& 23.39& 22.86& 22.52& 22.12& 21.86\\
J234804$-$302855  & 0.144$\pm$0.007   & 0.38$\pm$ 0.03  & 71   &  1.5$\pm$0.2 &23.93& 23.70& 23.88& 23.78& 24.38& 23.55& 23.22& 22.83&-\\
\hline \hline
\end{tabular}
\end{center}
\textsc{Note.} ---  Column 1 is the KiDS ID, in the form of hh-mm-ss and deg-mm-ss. Columns 2-5 list the source parameters from the lensing model by assuming a S\'{e}rsic profile for the sources. From left to right are the effective radius $R_{\rm eff}$ in the unit of arcsec, the minor-to-major axis ratio $b/a$, the position angle $pa$ and the \rui{r-band S\'{e}rsic index $n_r$, respectively. The Sersic indexes are fixed to the r-band value when modeling other band images.} Columns 6-14 list the magnitudes of each source obtained from the ``flexible'' lensing model (see text for details). Errors in magnitudes are shown in Fig. \ref{fig:SED}. \rui{(*)These are the two post-blue nuggets have been studied in N+20 with MUSE spectroscopy and re-analysed here to check the consistency of the 9-band lensing approach.}
\end{table*}

\begin{figure*}
    \hspace{-0.2cm}
    \includegraphics[width=18.5cm]{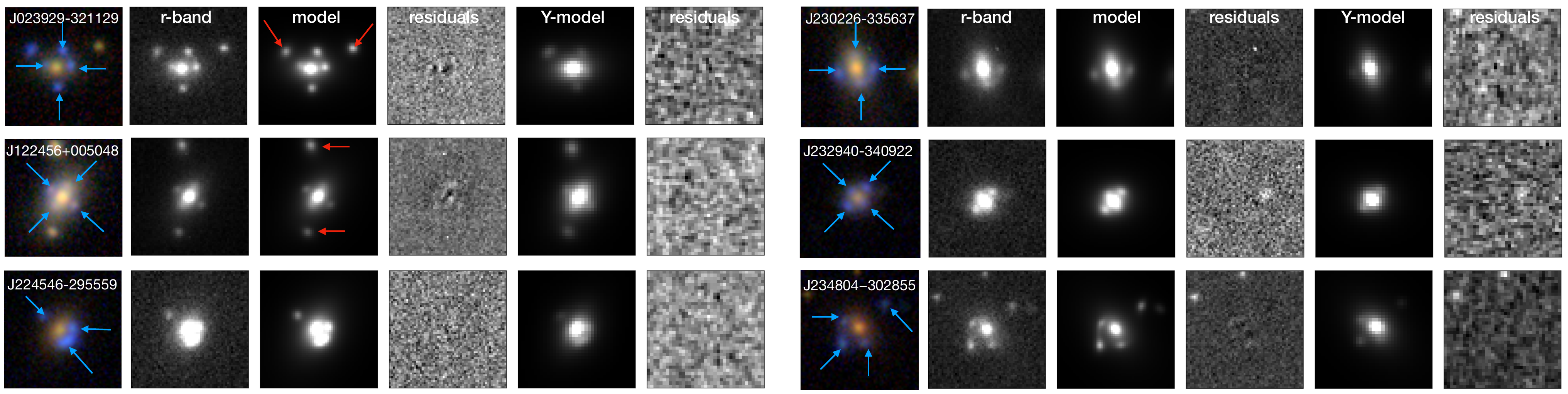}\vspace{3pt}
    \caption{Ray tracing model of the selected systems shown in Fig. \ref{fig:color_images}. From left to right: 1) the color image; 2) the $r$-band image; 3) the {\tt Lensed} $r$-band image model; 4) the $r$-band model-subtracted image; 5) the $Y$-band image model; 6) the $Y$-band model-subtracted image. The blue arrows point to the lensed images while the red arrows show other  projected galaxies in the field of view}.
    \label{fig:models}
\end{figure*}

\section{{Methods and model parameters}}
In this section we describe the %identification of the strongly lensed blue compact systems and the 
methods adopted to 1) perform the ray tracing models of the SL systems in optical and NIR images; 2) derive the structural parameters of the background sources; 3) use the ``unlensed'' multi-band photometry and colours to estimate the photo-$z$s and the stellar population parameters of the sources; 4) use the source decontaminated fluxes of the foreground lenses to obtain their photo-$z$s. Following N+20, we expect these systems to show 
%clear blue quadruply lensed images of 
high-$z$ ($z>1$), blue, ultra-compact background sources with sSFR compatible with quenched systems.
%, very close to Einstein crosses. 

\subsection{Lens Modeling}\label{modeling}
For the ray tracing modelling of the optical and NIR bands, we use {\tt Lensed} (\citealt{Tessore_2016MNRAS.463.3115T_lensed}), which implements an MCMC method to look for the best fitting parameters of the mass of the lens and the light of both the lens and background source. The lensing model is convolved with a point spread function (PSF) obtained by fitting one nearby star to the lens system with two Moffat profiles.
For the light distribution of the source, we assume a S{\'e}rsic profile (\citealt{1968adga.book.....S_Sersic}), %which is described as
%\begin{equation}
%I(x,y)=I_e \exp \{-b_n[(\frac{\sqrt{x^2+q^2y^2}}{R_{\rm eff}})^\frac{1}{n}-1]\},
%\end{equation}
where the free parameters are $n$, the S{\'e}rsic index, $R_{\rm eff}$, the effective radius, $x_0$ and $y_0$, the center coordinates, $q$, the axis ratio and $pa$, the position angle. The light of the lens is assumed to follow an elliptical de Vaucouleur profile (\citealt{deVaucouleurs1948AnAp...11..247D}), which corresponds to a special case of S{\'e}rsic profile with $n=4$.
%$b_n$ is determined by $n$ (\citealt{Ciotti_1999A&A...352..447C, MacArthur_2003ApJ...582..689M}). 
The total mass of the lens is modeled with a singular isothermal ellipsoid profile (\citealt{Kormann1994_SIE}), and its projected two-dimensional surface mass density profile is described by:
\begin{equation}
\Sigma(x, y)=\frac{1}{2}\Sigma_{\rm c}\, \sqrt{q}\, \theta_{\rm E}\, (x^2+q^2y^2)^{-1/2},
\end{equation}
where $\theta_{E}$ is the lensing strength, equivalent to the Einstein radius, and $q$ is the minor-to-major axis ratio of the isodensity contours.  $\Sigma_{\rm c}=c^2/(4\pi G) (D_{\rm s}/D_{\rm d} D_{\rm ds})$ is the critical density, where $D_{\rm s}$ and $D_{\rm d}$ are the angular diameter distances from the observer to the lens and the source, respectively, and $D_{\rm ds}$ is the distance between the lens and source. We also include the external shear, $\gamma_{\rm ext}$, which approximates the influence of the surrounding environment on the lensing potential. Finally, we have 16 parameters for the source, mass, and foreground light, all together.
%, where we just keep the center of the mass aligned with the one of the foreground light for simplicity.

As reference band for the ray tracing model, we use the $r-$band images, as they provide the best image quality with an average seeing of FWHM=$0.68''$. As shown in N+20, this, combined with the pixel scale of $\rm 0.2''/px$ of the camera (\citealt{Kuijken2019A&A...625A...2K}), is 
%especially important 
suitable to derive the structure parameters of the background sources, whose images are of the order of a few pixels (see Fig. \ref{fig:color_images}). \rui{
The best-fitting $r$-band parameters of the sources, obtainied with '{\tt Lensed}', are reported in Tab. \ref{tab:source_mod}.} The $\chi^2/dof$ obtained from the lensing modelling is always between 0.8 to 1.2, meaning that the 6 SL systems are accurately modelled. The intrinsic $r-$band AB magnitudes of the sources are between 23 and 24.7 and the effective radii, $R_{\rm eff}$, are between $0.05''$ and $0.15''$. These galaxies, faint in brightness and very compact in size, are hard to be directly resolved, even by space telescopes like HST. The S{\'e}rsic index of the sources is between $0.8<n<3.5$, indicating that some of them are likely to be disk galaxies ($n<2$), while some others have $n$-index closer to spheroids ($n>2$). We remark that, for the two previously studied systems in N+20, where a different ray tracing tool was adopted, we have found larger $n$ values (but yet all consistent with disk systems, i.e. $n\lsim1$), while the $R_{\rm eff}$ are consistent within 20\%. This shows that the size is a rather robust parameter, while the S\'ersic index is more uncertain.
%To collect the multi-band photometry necessary for stellar population analysis, we need to apply 
%The same lens modelling process is, then, applied to the other three optical bands ($ugi$) from KiDS and five NIR bands ($ZYJHKs$) from VIKING. As mentioned, the image quality of these eight bands is usually worse than the quality of the $r$-bands images, mainly because of the poorer seeing (which is set to $<0.7''$ only for $r$-band, according to the KiDS and VIKING survey strategies), but also for the lower SNR, especially in NIR (see \S\ref{target_selection}).

For all other three optical bands ($ugi$) from KiDS and five NIR bands ($ZYJHKs$) from VIKING, we use the same lens modelling set-up as for the $r-$band. However, the image quality of these eight bands is usually worse than the quality of the $r$-bands images, mainly because of the poorer seeing (which is set to $<0.7''$ only for $r$-band, according to the KiDS and VIKING survey strategies), but also for the lower SNR, especially in NIR (see \S\ref{target_selection}).
This requires some strategies to optimize the fidelity of the multi-band models. 

We start with modeling the 6 systems 
%with a rather ``flexible'' approach, where we 
by maximizing the number of free parameters in all bands\footnote{We notice that 3 systems have almost no signals from both the foreground and background galaxies in the {\it u}-band and 3 systems have very faint lensed images in the Ks-band, making the lens modelling in these bands impossible to perform. }. 
In this ``flexible'' prior approach, we fix the values of a few model parameters to those obtained in the $r$-band images. First, 
%we fix 
the position angle $pa$, and the axis ratio $q$ for the lens mass and the external shear. This is reasonable, as the mass model is independent of the band and $r$-band is expected to give the best constraints. We also check that the Einstein radius computed from each of the other bands, is consistent within the errors with the one found on $r$-band. This leaves some freedom to the mass model to incorporate the effect of data noise in the different bands. 
%Similarly, the external shear is also fixed.
Second, we also fix the axis ratio $q$ and S{\'e}rsic index for the background sources. 
This is also reasonable, as the KiDS and VIKING bands will correspond to NUV and optical rest frame bands if these are placed at $z>1$, and there is little evidence of strong color gradients at these wavelengths in either compact star-forming (e.g., \citealt{2016ApJ...822L..25Liu_UV-opt-grad})
or quiescent galaxies (e.g., \citealt{Whitaker2012ApJ...745..179W}).
%Finally, we notice that 3 systems have almost no signals from both the foreground and background galaxies in the u-band and 3 systems have very faint lensed images in the Ks-band, making the lens modelling in these bands impossible to perform. 
%Therefore, we abandon the modelling of these bands. 
%To increase the accuracy of the (poor image quality) NIR models we used the $r-$band as initial guess, but letting still full freedom to the model to converge to different parameters.

The final modelled magnitudes of the source of each lens, using this approach, are listed in Tab. \ref{tab:source_mod}, while in Fig.  \ref{fig:models} we show the ray-tracing models of the $r$-band, as representative of optical bands, and the $Y$-band, as representative of NIR bands.
%, are shown in Fig. \ref{fig:models}.
As a sanity check, we have also performed another round of models, where this time we have minimized the free parameters among the 9-bands. In particular, for the $r-$band we used the same approach as above, while for other bands, we used the $r-$band parameters to fix the effective radius $R_{eff}$ of the foreground lens, the $\theta_{E}$, $q$ and $pa$ of the masses, the external shear, as well as the $R_{eff}$, $n$, $q$, $pa$ of the sources. Although these ``frozen'' priors are based on stronger assumptions of the ``flexible'' ones, they reduce the freedom of the ray-tracing model to find arbitrary solutions that might impact the main parameters we are interested in, i.e. the source magnitudes. The results of the frozen prior models are reported in Appendix \ref{app:mod_compar}, where we show that the source magnitudes are not significantly affected by the two extreme approaches and discuss the comparison with the flexible prior results in detail.

\begin{table*}[htbp]
\begin{center}
\scriptsize
\label{tab:stellar_population}
\caption{\label{tab:stellar_population} Stellar population properties of the blue nuggets as in Table 1.}
\begin{tabular}{c c c c c c c c c c c c}
\hline \hline
$\rm z_{photo}^{lens}$ & $\rm z_{photo}^{src}$  &  $\rm R_{\rm eff}$ & $\rm \log M_{*Le}$& $\rm \log M_{*Ci}$ &  $\rm log(\rm sSFR)_{\rm Le}$ & $\rm log(\rm sSFR)_{\rm Ci}$ & $\rm age_{\rm Le}$ & $\rm age_{\rm Ci}$ &$\rm \Sigma_{\rm 1Le}$&$\rm \Sigma_{\rm 1Ci}$\\
& &(kpc)& ($\rm M_\odot$) &($\rm M_\odot$) &$\rm (Gyr^{-1}$)  &  $\rm (Gyr^{-1}$)  & (Gyr) & (Gyr) &$\rm M_\odot/kpc^2$&$\rm M_\odot/kpc^2$\\
\hline
 $0.46_{-0.09}^{+0.07}$& $2.08_{-0.15}^{+0.07}$  & 0.52$\pm$0.02  &$10.05_{-0.06}^{+0.19}$  & 10.32 & 0.43 &0.09& 0.5 & $2.0$ &9.40 &9.66 \\
$0.31_{-0.08}^{+0.22}$& $1.01_{-0.05}^{+0.06}$  & 0.46$\pm$0.02  &$10.16_{-0.21}^{+0.02}$  & 10.10 & -0.94 & -0.84& 4.5 & 4.0& 9.61 & 9.55 \\
$0.51_{-0.05}^{+0.04}$& $1.26_{-0.05}^{+0.08}$  & 0.45$\pm$0.05  &$9.98_{-0.25}^{+0.08}$  & 9.94 & -0.30 & 0.13 & 2.0 & 2.0&9.35 &9.30\\
$0.47_{-0.11}^{+0.04}$ & $2.14_{-0.11}^{+0.11}$  & 1.28$\pm$0.07  &$10.49_{-0.23}^{+0.09}$  & 10.41 & 0.01 & 0.40&1.0 &1.0 &9.62 &9.54 \\
$0.63_{-0.07}^{+0.09}$ & $1.62_{-0.07}^{+0.06}$  & 0.61$\pm$0.09  &$10.44_{-0.25}^{+0.01}$  & 10.44 & -0.36 & -0.30& 3.0 & 4.0& 9.76&9.76 \\
$0.49_{-0.04}^{+0.20}$ & $2.22_{-0.14}^{+0.06}$  & 1.19$\pm$0.06 &$10.23_{-0.21}^{+0.01}$ &10.18 & -0.36 &-0.20 & 2.5 & 3.0& 9.36&9.31\\
\hline \hline
\end{tabular}
\end{center}
\textsc{Note.} --- Column 1 shows the best-estimated lens photo-$z$s ($\rm z_{photo}^{lens}$) from {\tt Lephare}. Column 2 shows the corresponding source photo-$z$s ($\rm z_{photo}^{src}$). Column 3 shows the effective radius $R_{\rm eff}$ in the unit of Kpc. Columns 4-11 list the main stellar population parameters from both {\tt Lephare} and {\tt Cigale} (Le and Ci subfixes, respectively). From left to right are the stellar mass $\rm \log (M_{*}/M_{\odot})$, specific star formation rate $\rm \log (sSFR)$, the age, and the surface mass densities within the central 1kpc $\Sigma_1$, respectively. (*) Post-blue nuggets were studied in N+20 with no 9-band strong lensing analysis.
\end{table*}

\begin{figure*}
    \centering
    \includegraphics[width=17cm]{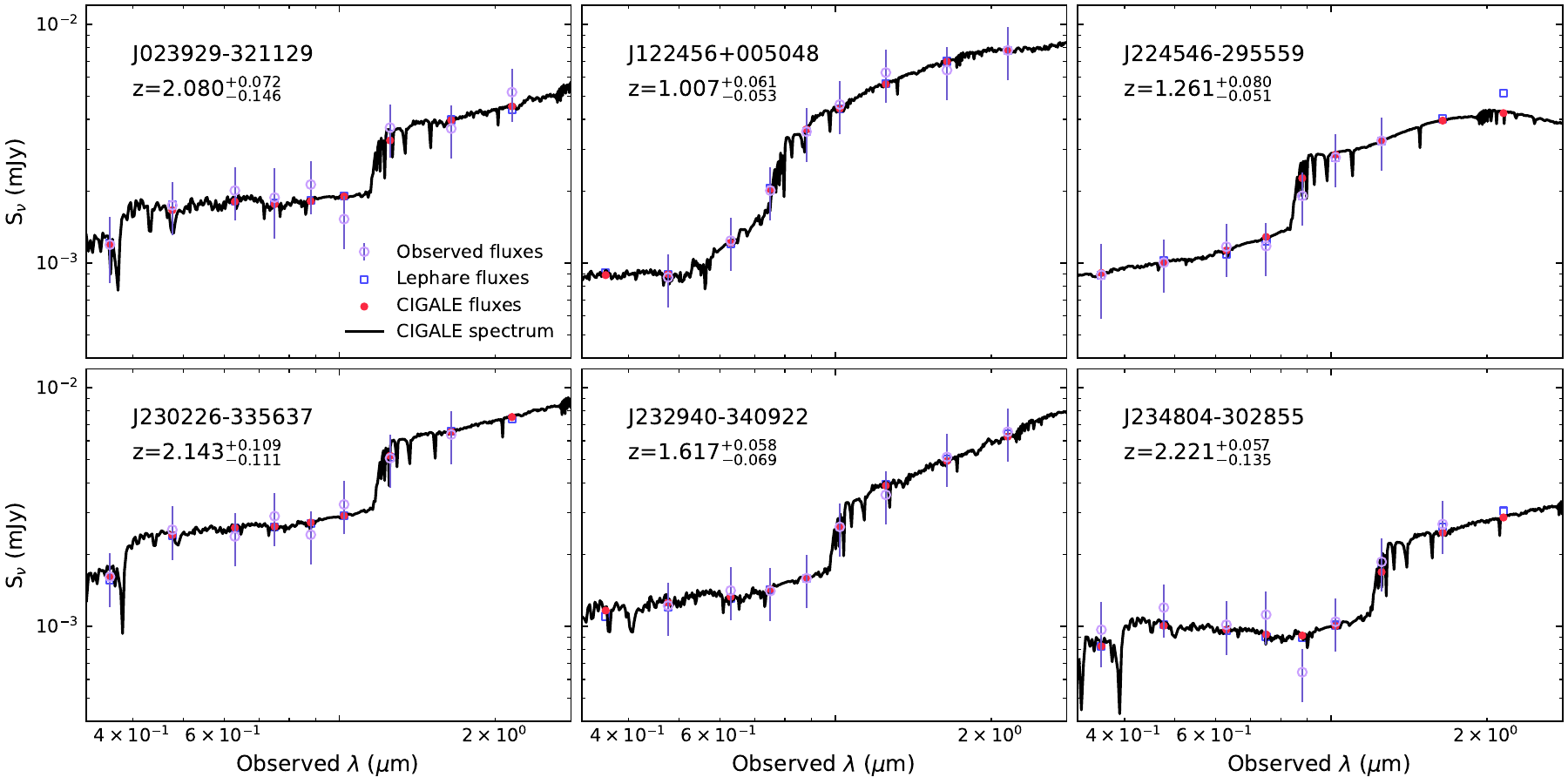}\vspace{3pt}
    \caption{The SED fitting results from {\tt Lephare} and {\tt Cigale} for the 6 source galaxies. In each panel, we show the observed multi-band flux from the lensing model (purple circles with error bars), the template flux from {\tt Lephare} (blue squares), and the template spectrum from {\tt Cigale} (black solid line) and its corresponding template flux (red dots). The 4000\ $\rm \AA$ break can be clearly seen in the spectra and the template flux, and their locations are consistent with the fitted photo-$z$s.}
    \label{fig:SED}
\end{figure*}

\subsection{Photo-$z$s and stellar population parameters of the sources from SED fitting}\label{sec:stellar}

Source redshifts are essential for two main reasons: 1) to confirm the high-$z$ nature of the multiple images and 2) to convert the angular sizes in arcsec into the physical scales in kpc to be used for the scaling relations.
As we have spectroscopic redshifts (spec-zs) for two systems only \rui{(i.e. the two systems previously analysed by N+20)}, we use the SED fitting of the 9-band photometry of the sources to determine the photo-$z$s. This approach has been found to be reasonably accurate for normal lenses using optical imaging from ground-based observations (see a.g. \citealt{Langeroodi2023A&A...669A.154L}). In our case, the unique availability of the NIR and the higher lensing model accuracy allowed by pseudo-cross lensing configurations (\citealt{2022ApJ...935...49Gu_GIGA}), suggesting that we can reach possibly a similar or better accuracy. \rui{As a sanity check for our approach we can also use the two N+20 lenses, for which we have the spec-zs for both the sources and deflectors, hence providing a direct validation of our photo-z estimates.}

\rui{The SED fitting is performed with {\tt Lephare}
(\citealt{Ilbert2006A&A...457..841I_lephare}) using as input the multi-band photometry of the pBNs derives with the lensing analysis as in Tab. \ref{tab:source_mod}.} Here, we adopt the BC03 (\citealt{BC03_2003MNRAS.344.1000B}) stellar population synthesis models, a Chabrier (\citealt{Chabrier2003PASP..115..763C}) stellar Initial Mass Function and exponentially decaying star formation history. As we expect these systems to be compact, poorly star-forming, old systems, we assume solar metallicity (as found for previous lensed pBNs in N+20\footnote{However, the assumption of solar metallicity is rather conservative for high-$z$ galaxies. Using more realistic lower metallicities would produce higher ages (\citealt{Worthey1994ApJS...95..107W} \citealt{Trager2009MNRAS.395..608T}), hence reinforcing the evidence of old stellar populations in the source galaxies.}), while all other parameters, such as the e-folding time, age of the main stellar population and internal extinction, E(B-V), are free to vary. 
\rui{For each of the six pBNs,} we let the redshift vary in the range of (0, 4), and require {\tt Lephare} to output the best-fitting parameters, e.g. photo-$z$s, stellar mass, age, SFRs, and their corresponding errors. The resulting photo-$z$s, as well as other stellar population parameters, are shown in Tab. \ref{tab:stellar_population}.
%\rui{, for both the model set-ups described in the previous section}.
\rui{To check the robustness of our photo-$z$ estimates, we verify the consistency of the results from the ``flexible" and ``frozen" models (see \S \ref{modeling}).
%and further compare the photo-$z$ and spec-$z$s of the two N+20 spectroscopically confirmed lenses.
The comparison between the two models (see also Appendix \ref{app:mod_compar}) shows that the photo-$z$s from different models are consistent within the typical errors expected for high-$z$ galaxy photo-$z$s from ground-based observations (e.g. \citealt{2020A&A...642A.200V,2021inas.book..245A,2022A&A...666A..85L}). The flexible result also shows that the N+20 objects have photo-$z$s of $1.01_{-0.05}^{+0.06}$ and $1.62_{-0.07}^{+0.06}$, respectively, which are fully consistent with their spec-$z$s (i.e. $1.101$ and $1.589$, respectively).}

%The former check (see also Appendix \ref{app:mod_compar}) shows that the photo-$z$s from the two models are consistent within the typical errors expected for high-$z$ galaxy photo-$z$ from ground-based observations (e.g. \citealt{2020A&A...642A.200V,2021inas.book..245A,2022A&A...666A..85L}). The latter check shows that N+20 sample has photo-$z$s of $1.01_{-0.05}^{+0.06}$ and $1.62_{-0.07}^{+0.06}$, respectively, which are fully consistent with their spec-$z$s of $1.101$ and $1.589$.

The source redshifts of the 6 SL systems cover the range of $1.0\lesssim z\lesssim 2.3$.
%with half of them at $z\sim2-2.3$, and the other half at $z\sim1.0-1.6$. 
In Fig. \ref{fig:SED}, we show the observed fluxes and the best-fitted SED templates for the sources for the ``flexible'' model only.
The best fit of the multi-band photometry is remarkably good in all cases. In particular, the NIR bands are essential to closely map the rest-frame 4000 \AA\ break, which is essential in the photo-$z$ determination. 

As a consistency check, to make sure that the parameters obtained by {\tt Lephare} are tool-independent, we repeat the same analysis with {\tt Cigale} (v2018.0, \citealt{Boquien2019A&A...622A.103B_Cigale}).
{Since this latter is not designed to predict photo-$z$s, we fix the redshift of each source to the photo-$z$s from {\tt Lephare}. \ruili{The resulting stellar mass and age of the sources (see Tab. \ref{tab:stellar_population}) are consistent with {\tt Lephare} estimates. 
We observed that the sSFRs derived from the {\tt Cigale} show values that are close to the ones of {\tt Lephare}, except for J023929$-$321129, and assuming typical errors on sSFR from {\tt Lephare} ($\sim 0.4$ dex, see \citealt{Xie2023arXiv230704120X}) they are all consistent within $1\sigma$.}  
%are marginally higher than those obtained through the {\tt Lephare}. However, this discrepancy is very slight and can be negligible. Notably, the galaxy KiDSJ023929-321129 stands out as an exception, which turns out to be more massive and old system, with a lower $\rm log(sSFRs)$ more compatible with those inferred for the other five systems.
%We found the sSFRs from {\tt Cigale} are slight higher than that from {\tt Lephare}, but can be ignored. KiDSJ023929$-$321129 is a exception, which turns out to be more massive and old, with a lower $\rm log(sSFRs)$ 
%is smaller with {\it Cigale}, 
%but yet consistent with a rather small star formation rate 
%more compatible with those inferred for the other five systems.
%yet compatible with the one obtained for all other systems, all consistent with very low sSFRs (see also below). 
As {\tt Cigale} does not provide uncertainties on the best-estimated parameters, we will use {\tt Lephare} results as a reference for stellar population parameters in the following.} We finally remark that all stellar population estimates are also consistent with the ones obtained with the ``frozen'' model magnitudes, within the typical statistical errors of stellar populations, as discussed in Appendix \ref{app:mod_compar}.

Finally, in Table \ref{tab:stellar_population}, we also report the photo-$z$s of the foreground lenses, that have been derived following the same procedure for the source photo-$zs$. The details of this part of the analysis are beyond the purpose of this paper and will be the focus of an upcoming ray-tracing analysis.
%(Li et al., in preparation).
Here we just remark that all the systems are consistent with being lensing events, 
%Here we just remark that they are all consistent with all systems being lensing events,
as the redshifts of the central galaxies are lower than the one of the background sources. \rui{The photo-$zs$ of the lenses of the two spectroscopic systems from N+20 are $0.31_{-0.08}^{+0.22}$ and $0.63_{-0.07}^{+0.09}$, respectively,
both are higher than their spec-$z$s of $0.237$ and $0.381$. The former aligns well with the spec-zs, within approximately one standard deviation (
$1\sigma$) of error. However, the latter exhibits a discrepancy greater than 3$\sigma$. We intend to reserve further refinement of these foreground redshifts in future researches.
%The former is consistent within $\sim 1\sigma$ error while the later is worse, with an error exceed 3$\sigma$. We left further refine on these foreground redshift to future works.    %which is rather acceptable for reasonably accurate photo-$zs$. 
} 

%both are higher than their spec-$z$s of $0.237$ and $0.381$, but consistent within $\sim2\sigma$ in the worse case, which is rather acceptable for reasonably accurate photo-$zs$. 

%from here

\section{Results and Discussion}\label{discussion} 
In this section, we concentrate on the results obtained with the flexible model, while the ones from the frozen approach are discussed in Appendix \ref{app:mod_compar}. 
We start by using the photo-$z$s derived in Tab. 2 to convert the effective radii $\rm R_{\rm eff}$ into the physical scale. We find that all sources have $\rm R_{\rm eff}$ smaller than 1.5 kpc, while stellar masses are in the range of $ 9.9< \rm \log (M_*/M_{\odot})<10.5$, as reported in Tab. \ref{tab:stellar_population}. In the same table, the surface mass densities within 1 kpc are listed and are larger than $\rm \log (\Sigma_1/M_{\odot} \rm kpc^2)\sim 9.3$. These numbers make the sources consistent with typical parameters of blue nuggets (e.g., \citealt{Zolotov2015MNRAS.450.2327Z}, \citealt{Tacchella2016MNRAS.458..242T}, \citealt{Barro2017ApJ...840...47B}). The ages of these galaxies are generally older than 1.0 Gyr, except for J023929-321129, which has an age of only 0.5 Gyrs. 
%\rui{We acknowledge that the aforementioned relative old age is predicated on the assumption of solar metallicity, and the associated uncertainties may be amplified if this assumption were to be abandoned. Nevertheless, these ages retain their significance, as comparable aged stellar populations were identified through a comparable 9-band spectral energy distribution analysis in N+20, and were further supported by integral field spectroscopy that revealed absorption-dominated spectral characteristics characteristic of a mature stellar population.}
%\rui{Here we note that this relative old age is obtained from the assumption of solar metallicity, the uncertainties would probably increase if this assumption was dropped. However, these ages are still meaningful, because relatively old stellar populations were also found from a similar 9-band SED analysis in N+20 and corroborated by the integral field spectroscopy showing absorption-dominated spectral features typical of an evolved stellar population.}
%The two N+20 systems remain the oldest ones in the sample with the new SL-based 9-band photometry showing ages of 4.5 (J122456$+$005048) and 3 (J232940$-$340922) Gyr, i.e. having less than 25\% error compared with previously found in N+20 (3.8 and 4.5Gyr, respectively). 
\rui{The two N+20 systems remain the oldest ones in the sample with the new SL-based 9-band photometry, showing ages of 4.5 (J122456$+$005048) and 3 (J232940$-$340922) Gyr. This exhibits a less than 35\% discrepancy when compared to the ages previously reported in the N+20 study, which were 3.8 Gyr and 4.5 Gyr, respectively.}
All systems show low sSFRs, i.e. $\log \rm (sSFR/Gyr) \lesssim 0.4$, which are consistent with the definition of quiescent systems (see \citealt{2015MNRAS.453.2540J}, \citealt{Zolotov2015MNRAS.450.2327Z}, \citealt{Barro2017ApJ...840...47B}, \citealt{Huertas-Company2018ApJ...858..114H}). In conclusion, in line with the findings presented in N+20, these systems are all fully compatible with being pBNs.
%\rui{Five systems exhibit markedly low specific star formation rates, with $\log \rm (sSFR) \lesssim 0.2\ M_\odot Gyr^{-1}$, consistently with the definition of quiescent systems (see \citealt{2015MNRAS.453.2540J}, \citealt{Zolotov2015MNRAS.450.2327Z}, \citealt{Barro2017ApJ...840...47B}, \citealt{Huertas-Company2018ApJ...858..114H}), while KiDS J023929$-$321129, 
%while having a slightly higher value of $\log \rm (sSFR) = 0.43\ M_\odot Gyr^{-1}$, 
%is just above the upper limit for quenching systems ($\log \rm \  sSFR \lesssim 0.3\ M_\odot Gyr^{-1}$, e.g. \citealt{Zolotov2015MNRAS.450.2327Z}). In summary, these systems are consistent with being pBNs, which is in line with the outcomes reported in N+20.}

\begin{figure*}
\hspace{-0.5cm}
\includegraphics[width=18.5cm]{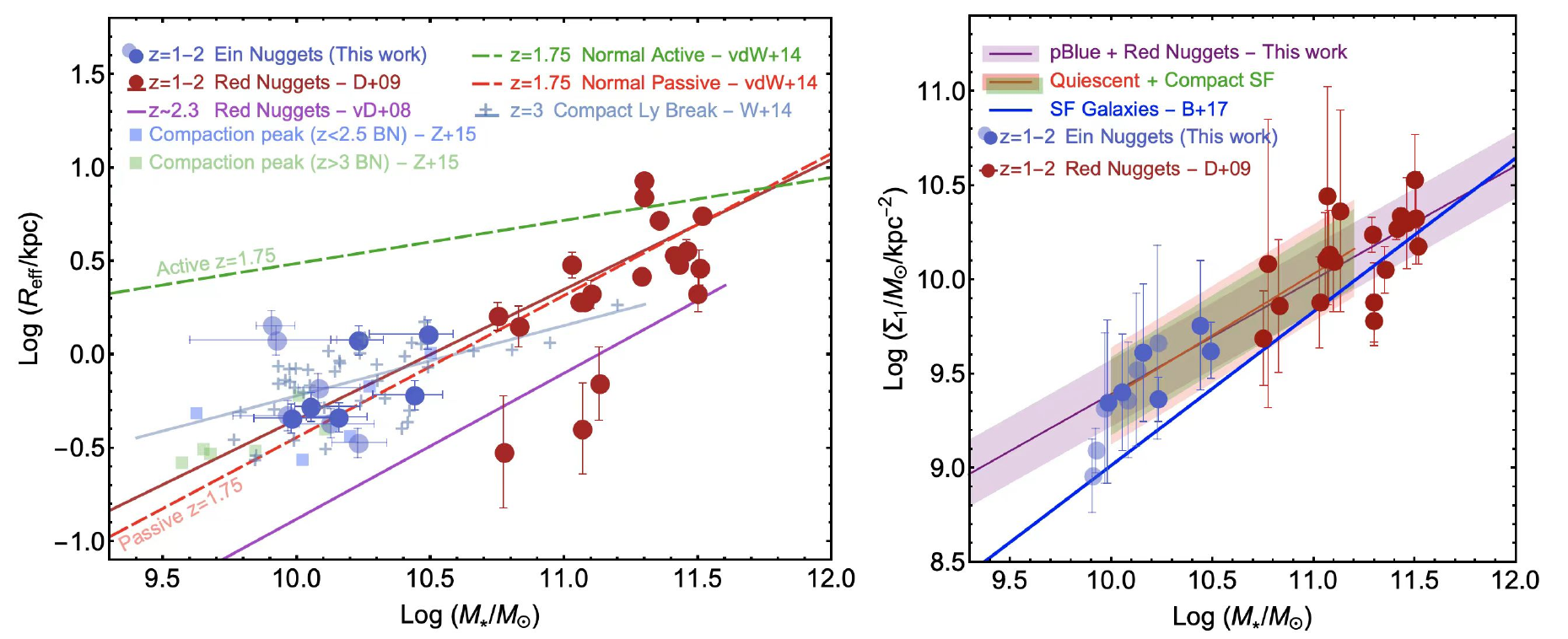}\vspace{3pt}
\caption{Left: size-mass relation of the pBNs (blue dots) and other galaxies from literature, i.e., 1) red nuggets at $1<z<2$ and their linear fit (red dots and red solid line, \citealt{Damjanov2009ApJ...695..101D}, D+09), and linear fit to the red nuggets at $z\sim2.3$ (purple line, \citealt{vanDokkum2008ApJ...677L...5V}, vD+08); 2) the compaction peak of blue nuggets at $z<2.5$ (green squares) and $z>3$ (blue squares) in simulations (\citealt{Zolotov2015MNRAS.450.2327Z}, Z+15); 3) the compact LBGs at $z\sim3$ and their linear fit (blue crosses and blue solid line); 4) linear fit to the measurements of normal passive (red dash lines) and active (green dash lines) galaxies at $z=$1.75 (\citealt{vdW2014ApJ...788...28V}, vdW+14). All of the \ruili{stellar masses of the pBNs} have been corrected to Chabriar-IMF.  Right: $\Sigma_1$-$\rm M_*$ relation of the pBNs (blue dots) and red nuggets (red dots) at $1<z<2$ from \cite{Damjanov2009ApJ...695..101D}. The solid purple line is the best-fit relation, while the purple shaded
area represent the $1\sigma$ error of the fitted relation. \ruili{We also plot the same relationship for samples of quiescent galaxies (red shaded area), compact star-forming galaxies (green shaded area) and the normal star-forming galaxies (blue line) from \citet[][B+17]{Barro2017ApJ...840...47B}.}
In both plots, it is clear that the pBNs follow the trend of the passive galaxies and Red Nuggets, rather than standard active galaxies at the same redshift.}
\label{fig:mass_reff}
\end{figure*}
%For instance, \citet{Huertas-Company2018ApJ...858..114H} shows that pBNs occurr at an average $\rm log\ sSFR$ ($\sim 0 \ M_\odot Gyr^{-1}$) higher than predicted in simulations ($\sim -0.5 \ M_\odot Gyr^{-1}$), while \citet{Johnston2015MNRAS.453.2540J} find even higher $\rm log\ sSFR$ 
Further supports for this conclusion comes from other similar results obtained in literature: 1) in terms of the SFR--$\rm M_*$ \footnote{\ruili{Here we switch to SFR instead of using sSFR, to direct compare our results with the literature.}} relation \citet[][see their Fig. 6]{Johnston2015MNRAS.453.2540J} show that for galaxies with masses of $\rm 9.9\lesssim log(M/M_{\odot}) \lesssim 10.5$, the transition place between SFGs and passive galaxies occurs at $\rm 0.5 \lesssim  log (SFR/M_{\odot} yr^{-1}) \lesssim 1.5$, compatible to what found for our pBN sample, \ruili{
%reported in Table \ref{tab:stellar_population}
$\rm 0.2 \lesssim  log (SFR/M_{\odot} yr^{-1}) \lesssim 1.5$ (calculated from Tab. \ref{tab:stellar_population} \footnote{$\rm log (SFR/M_{\odot} yr^{-1})=log (sSFR/Gyr)+log(M_*/M_{\odot})-9$ }});
%$\rm 0.2\lesssim log(SFR/M_{\odot} yr^{-1})\lesssim 1.6$\footnote{This can be seen from Table \ref{tab:stellar_population} where $\rm log (SFR/M_{\odot}yr^{-1}) = \log (sSFR) + \log M_* -9$.};
2) in terms of the sSFR--$\Sigma_1$ relation, \cite{Huertas-Company2018ApJ...858..114H} shows that the pBNs at $1.4<z<2.2$, and with a central mass density of $\rm log(\Sigma_1/\ M_{\odot} \mathrm{kpc}^{-2})\gsim 9.0$, have $\rm -1.5\lesssim log (sSFR/Gyr)\lesssim 0$.
This latter study also shows that the density of pBNs in the above mass and redshift ranges is higher than the density of blue nuggets. For this reason, and for the ``selection effect'' from the faint NIR photometry, making us little sensitive to strong star-forming systems at $z>1$, as discussed in \S\ref{target_selection}, we tend to select blue (due to the depth in optical bands), compact (due to the strong lensing effect) \ruili{systems with low star formation rate}, all conspiring toward a selection of pBNs. 

\begin{figure}
\hspace{-0.5cm}
\includegraphics[width=8.5cm,height=7.7cm]{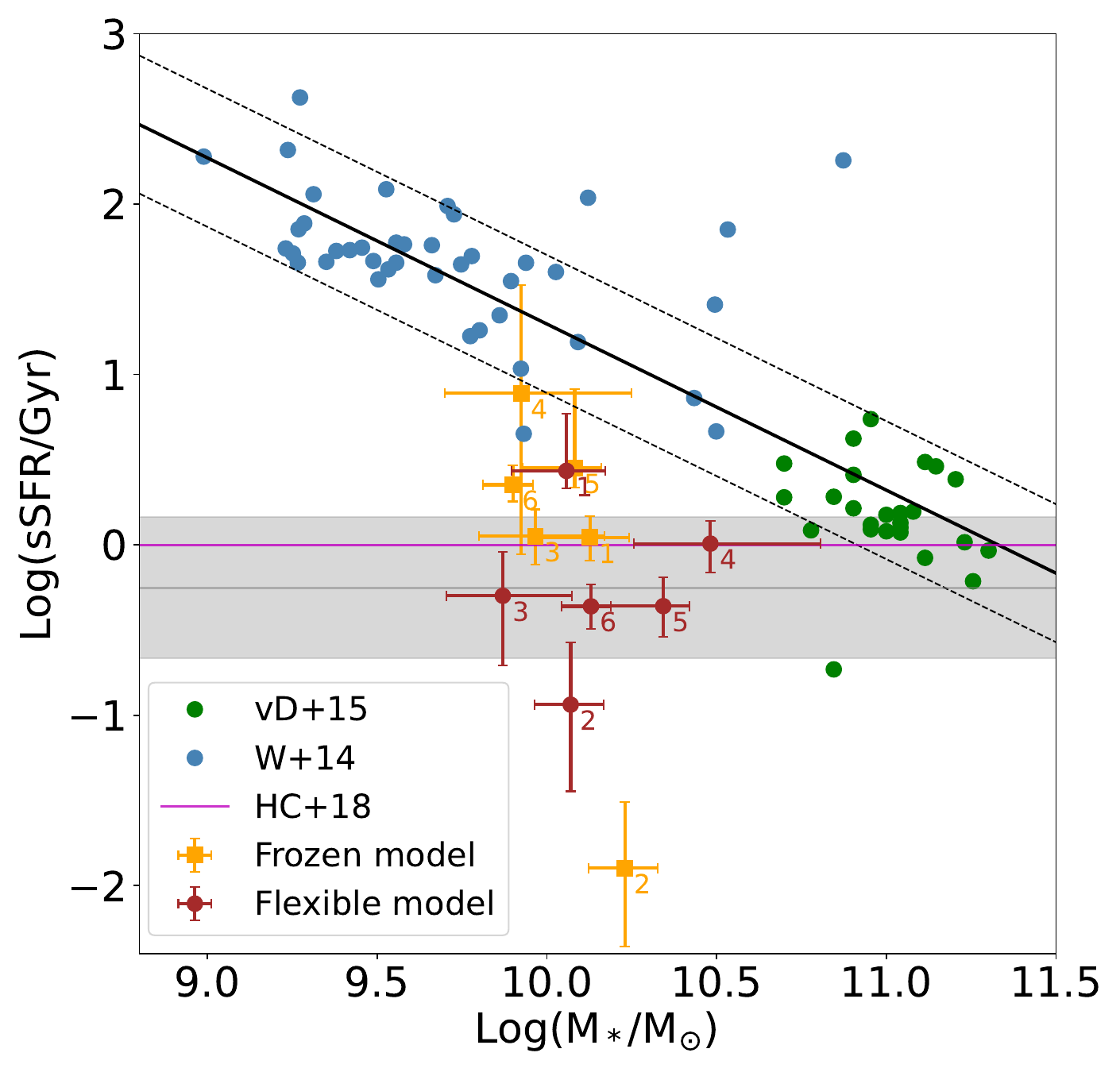}\vspace{3pt}
\caption{\rui{The sSFR--$\rm M_{*}$ relation for the 6 pBNs from both the flexible model (red dots with error bars) and the frozen model (orange squares with error bars). The grey horizontal line represents the mean $\rm log(sSFR)$, with the shaded area indicating the $1\sigma$ confidence interval. The numbers 1-6 represent the six nuggets in the order listed in Table \ref{tab:source_mod}. For comparative purposes, we have also plotted the compact massive SFGs (green dots) within the redshift range of $2.0<z<2.5$ \citet[][vD+15]{vanDokkum2015ApJ...813...23V}, the compact LBGs (blue dots) at $z\sim3$ (\citealt{Williams2014ApJ...780....1W}, W+14) and the up limit of the $\rm log(sSFR)$ for pBNs (magenta line) in CANDELS at redshift $1.4<z<2.0$ (\citealt{Huertas-Company2018ApJ...858..114H}, HC+18). All mass estimates have been rescaled to be compatible with a Chabrier IMF. The compact massive SFGs and the LBGs are distributed along a single relation (solid black line), with the $1\sigma$ confidence interval $\sim$0.4 (dashed black lines). whereas the pBNs are situated systematically below this line, suggesting a transition out of the star-forming phase.}}
\label{fig:mass_ssfr}
\end{figure}

The size-mass relation for the 6 pBNs is plotted in the left panel of Fig. \ref{fig:mass_reff} (blue solid circles).
%The size-mass relation of the 6 pBNs (blue solid circles) is plotted in Fig. \ref{fig:mass_reff} (left panel). 
Here, we also show: 1) red nuggets at $z\sim1.5$ from \citet[][D+09, red solid circles and best fit linear line]{Damjanov2009ApJ...695..101D} and at $z\sim2.3$ from \citet[][vD+08, purple solid line]{vanDokkum2008ApJ...677L...5V}; 2) compact Ly$\alpha$ break galaxies (LBGs) at $z\sim3$ from \citet[][W+14, blue crosses]{Williams2014ApJ...780....1W}; 3) normal active galaxies (green dash lines) and passive galaxies (red dash lines) at z=1.75 from \citet[][vdW+14, \ruili{similar data can also be found in \citealt{Mowla2019ApJ...880...57M} and \citealt{Nadolny2021A&A...647A..89N}}]{vdW2014ApJ...788...28V}.
All stellar masses are corrected to a Chabrier-IMF, consistently with our choice.
From Fig. \ref{fig:mass_reff}, we see that our lensed pBNs are located away from the size-mass relationship of the SFGs (green dashed line). Rather, they mainly occupy the location of the 
compaction peak of blue nuggets at $z<2.5$ (blue boxes) and $z>3$ (green boxes) predicted by simulations (\citealt{Zolotov2015MNRAS.450.2327Z}, Z+15). This shows that the typical masses and sizes of the 6 systems are fully compatible with post-compacted galaxies.
The log-linear mass-size relation for the pBNs and red nuggets, $\rm log (R_{\rm eff})=\alpha log (M_*)$, gives a slope $\alpha\sim0.66$.
Compared to ``red nuggets'' and passive galaxies at similar redshift ($z=1.5$ and $z=1.75$ respectively), the lensed pBNs are generally aligned to the size-mass relations of both samples, while they all seem offset by $\sim0.5$dex with respect to the red nugget samples at $2<z<3$ (%$\alpha\sim0.3$,
\citealt{vanDokkum2008ApJ...677L...5V}), possibly due to the earlier formation of these higher-$z$ systems. Finally, the pBN sample looks aligned to the compact LBGs (blue crosses), which are believed to be blue nuggets with strong star formation, $\rm log (sSFR/Gyr)>0.5$ or $\rm log (SFR/M_{\odot} yr^{-1})>3$(\citealt{Williams2014ApJ...780....1W}), however, the slope of these latter is $0.38$ (light blue line in Fig. \ref{fig:mass_reff}), slightly flatter than that of the pBNs plus red nuggets. This confirms the scenario that pBNs share properties of both active and passive compact galaxies, because the quenching occurs after compaction in the pBN phase and the 6 pBN systems seems indeed in the transition from typical ``blue nugget'' systems to the ``red nuggets''.
The turning knee of the two size-mass relationships appears at the location of $\rm log (M_*/M_{\odot})\sim$10-10.5. This has been indicated as a characteristic mass scale for a structural transition of the red-blue nugget systems (see e.g. \citealt{Lapiner2023MNRAS.522.4515L}), mainly because of the dominance of the wet processes (gas inflows) at lower mass scales, and the dry processes (e.g. dry mergers) acting at larger mass scales. Finally, in the same Fig. \ref{fig:mass_reff}, we show, as lighter blue data, the results of the frozen model for comparison. We see that two of the systems deviates from the
%$R_{\rm eff}-M_*$ 
size-mass relation. If we take their sSFRs at face value (see Fig. \ref{fig:mass_ssfr}), these two systems turn out to be very close or superior to the \ruili{upper sSFRs (or SFR) limits of quiescent systems or pBNs} discussed in literature (e.g. \citealt{Johnston2015MNRAS.453.2540J}, \citealt{Huertas-Company2018ApJ...858..114H}, see above). This leaves us with the possibility that 2 of the 6 systems are still in the transition to the blue nugget phase, as also suggested by their $\Sigma_1$ (see below).

On the right panel of Fig. \ref{fig:mass_reff}, we show the density inside 1 kpc vs. stellar mass of the pBNs. Here, the alignment with red nuggets is even more evident than the size-mass relation on the left panel. The log-linear relation gives a slope of 0.65, fully consistent with the one found for the quiescent galaxies in the redshift range $1.4<z<2$ by \citet[][B+17, also reported in Fig. 4 -- right]{Barro2017ApJ...840...47B}. This latter work also shows that the compact SFGs follow the sequence of the quiescent galaxies, consistently with the scenario that compaction produces a fast core growth causing normal SFGs to move up from the main SFG sequence (see blue line reported in Fig. \ref{fig:mass_reff} -- right). In this respect, the two frozen models outliers seen above, occupy the region of the transition from the star forming systems to compact SFGs, suggesting that they might even still going through compaction. 
This indicates that the classification of these two systems as pBNs is the most insicure one and they might therefore need a spectroscopic follow-up.

\rui{Fig. \ref{fig:mass_ssfr} presents the $\rm log(sSFR)$ as a function of the stellar mass $\rm log(M_*)$ for the six pBNs, using both the  ``flexible'' model (red dots with error bars) and the ``frozen model'' (orange dots). The mean $\rm log (sSFR)$, calculated to be -0.253Gyr, is showed by the grey horizontal line. The $1\sigma$ confidence interval, determined by the standard error ($\sigma=0.42$), is reported as a shaded region. This standard error coincides with typical uncertainties ($\sim 0.4$ dex) in sSFR estimates derived from 9-band KiDS photometry and photo-$z$s (see \citealt{Xie2023arXiv230704120X}), hence showing that the dominating effect on the scatter comes from the SED procedure rather than the lensing/photometry errors. In the same figure, we also plot compact massive SFGs(green dots) within the redshift range $2.0<z<2.5$ from \citet[][vD+15]{vanDokkum2015ApJ...813...23V}, and the compact LBGs (blue dots) at $z\sim3$, from \citet[][W+14]{Williams2014ApJ...780....1W} for comparison, similarly to what showed in Fig. \ref{fig:mass_reff}. Both these SFG samples
%segregated in mass by construction, 
clearly sit on a common log--log linear trend, represented by black solid line, with a standard deviation of $\sigma=0.41$, represented by black dashed lines. 
%On the other hand, almost all pBNs are observed to lie $\sim 2\sigma$ away from this fitted line, with the ``flexible'' model being more robustly below the correlation of the star-forming sample, suggesting that their star-forming activities are indeed weaker than those of typical SFGs. 
On the other hand, all pBNs are observed to lie below the fitted line to the star-forming sample, suggesting that the star-forming activities of pBNs are indeed weaker than those of typical SFGs.
The magenta horizontal line indicates the upper-limit of the typical $\rm log(sSFR)$ found for pBNs in CANDELS (\citealt{Huertas-Company2018ApJ...858..114H}, HC+18) at redshift $1.4<z<2.0$. Most of our pBNs are consistent
%or below 
this line. Overall we tend to conclude that the final results are independent on the lensing priors (frozen vs. flexible) adopted.
%To ascertain the model independence of this conclusion, we also depict the same $\rm log(sSFR)-log(M_*)$ relationship as derived from the ``frozen'' model. Although the $\rm log(sSFR)$ are slightly higher than those of the ``flexible'' model, they still reside 1 standard deviation away from the star-forming galaxies, which substantiates the notion of a weak star-forming nature.
}

\section{conclusions}
We have collected 6 strong gravitational lens systems with blue point-like lensed images from the catalog of ``high-quality candidates'' from the KIDS Survey (\citealt{Petrillo2019MNRAS.484.3879P, Li2020ApJ...899...30L, Li2021ApJ...923...16L}). We have performed ray-tracing modelling in the 9-band imaging data and derived the intrinsic multi-band photometry of the sources. Via photo-$z$ and  stellar population analysis, we have confirmed that 1) the systems are genuine SL events; and 2) the background sources of these 6 SL systems are very likely pBNs because of their very small effective radii, large mass, and low sSFRs, although, 3) depending on the priors on the ray-tracing model (i.e. whether one leave almost multi-band parameters free or fixed to the highest quality $r$-band model), there is a possibility that  2 of the 6 systems are still in a transition phase.
In particular, from the perspective of the SFR--$\rm M_*$ and sSFR--$\rm \Sigma_1$ relations, the 6 pBNs are well consistent with observations of compact low sSFR objects in \cite{Johnston2015MNRAS.453.2540J} and 
%the post-BN sample in 
\cite{Huertas-Company2018ApJ...858..114H}. In terms of the $\rm R_{eff}$--$\rm M_*$, $\rm \Sigma_1$--$\rm M_*$ and $\rm sSFR$-$\rm M_{*}$ relations, these pBNs
do not follow
%stay far away from
the main sequence of SFGs but align well with the relation of the passive, compact red nuggets at similar redshifts. Also, they sit on the tail of the size mass relation of the compact LBGs
from \citet{Williams2014ApJ...780....1W}. However, the two systems showing up as outlier in the frozen priors' model, taken at face value, possess a SFR which is compatible with ongoing star formation and occupy the region of these scaling relations of transient objects from normal SFGs to the pBN systems.
Overall, considering the continuity of the size-mass relations of the compact SFGs and the red nuggets, the ``confirmed'' pBNs (especially looking at the flexible prior results) sit in a knee located at $\log (M_*/M_\odot)\sim 10.5$ consistently with the presence of a characteristic mass for wet compaction (\citealt{Lapiner2023MNRAS.522.4515L}). 

All these facts together indicate that the pBNs are in a transition phase between the blue and red nuggets, as all structural scaling relations are set to become red nuggets, but with some residual ongoing star formation. 
This latter is seen, for the two systems in \cite{Napolitano2020ApJ...904L..31N}, from weak emission lines coexisting however with stellar absorption lines typical of old stellar populations. 
Finally, taking the modeled S{\'e}rsic indexes of these 6 pBNs at face value, 
regardless of the adopted prior, 
%regardless the priors adopted
they are generally smaller ($n<3$) than the predictions from simulations ($n>3$) or observations of red nuggets (\citealt{Damjanov2009ApJ...695..101D}). Nevertheless, these values are compatible with the scenario that, during the pBN phase, i.e. after the gas is depleted from the centre leaving behind a compact passive nugget, a new extended rotation-supported gas disc/ring can develop (\citealt{Toft2017Natur.546..510T, Lapiner2023MNRAS.522.4515L}). 
The only way to obtain a more detailed model and shed light on this aspect would be to obtain higher quality images.
%However, more firm results on the model details will possibly need higher image quality data.

This work is based on a small number of objects (possibly all we can expect, according to N+20) found
%that had a high chance to turn to be real SLs 
in the KiDS footprint, but other candidates have been identified in the footprint of other surveys like HSC (\citealt{Jaelani2020MNRAS.494.3156J}) and DES (\citealt{Jacobs2019ApJS..243...17J}). We plan to extend this analysis to other systems soon, and possibly add spectroscopic data in order to further observationally support the understanding of the pBN phase
%of galaxy evolution, 
which is crucial in the morphological transformation of galaxies in their early phase of evolution.

\section*{Acknowledgements}
Rui Li and Ran Li acknowledge the support of the National Nature Science Foundation of China (Nos 12203050 11988101,11773032,12022306), the science research grants from the China Manned Space Project (CMS-CSST-2021-A01) and the support from K.C.Wong Education Foundation.
NRN acknowledges financial support from the Research Fund for International Scholars of the National Science Foundation of China, grant n. 12150710511 (BluENeSS). CT acknowledges the INAF grant 2022 LEMON. Xiaotong Guo is supported by \emph{Anhui Provincial Natural Science Foundation} project number 2308085QA33.

\bibliography{pBNs}{}
\bibliographystyle{aasjournal}

\begin{appendix}
\setcounter {table} {0} \renewcommand {\thetable} {\Alph {section}\arabic {table}}
\setcounter {figure} {0} \renewcommand {\thefigure} {\Alph {section}\arabic {figure}}

\section{Multi-band ray-tracing frozen model results}
\label{app:mod_compar}
\begin{figure*}
%\hspace{-0.5cm}
\centering
\includegraphics[width=14cm]{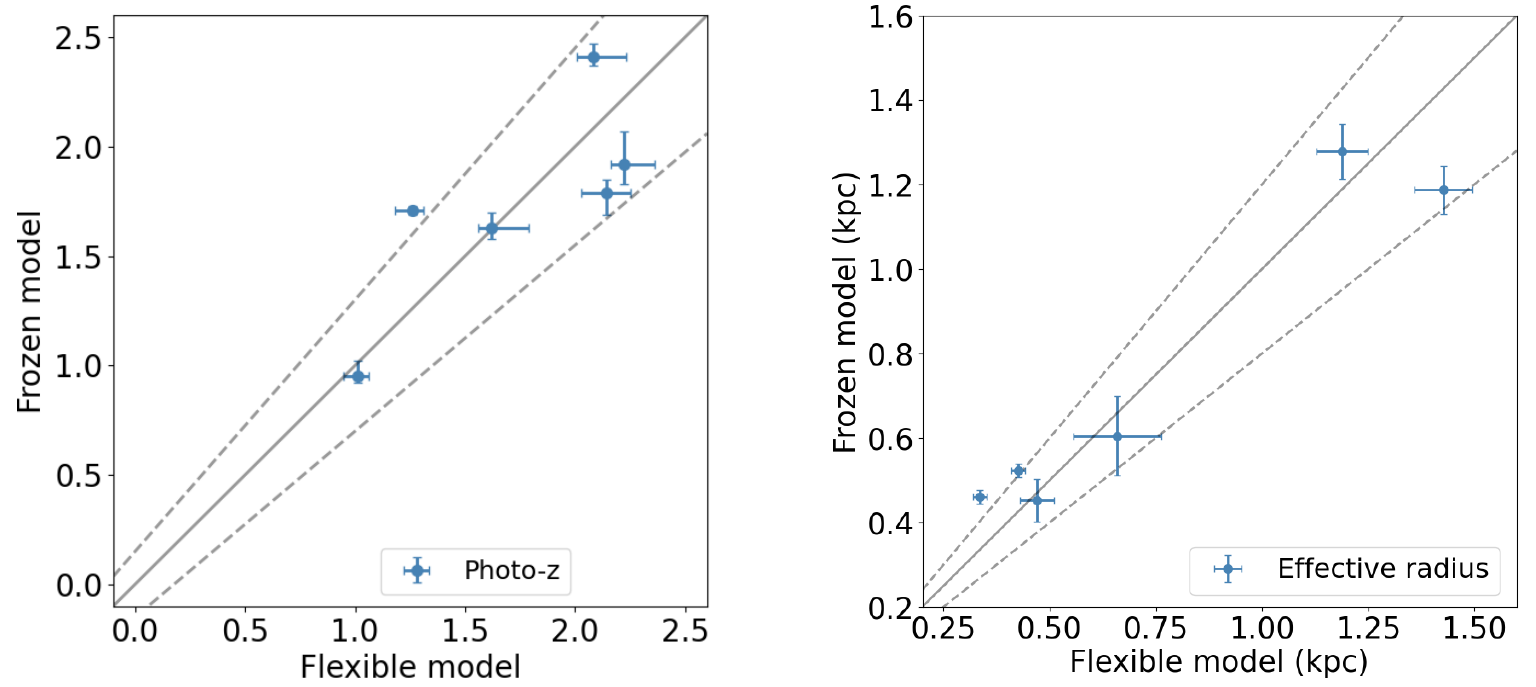}\vspace{3pt}
\caption{Comparison of relevant parameters obtained from the frozen model and flexible model. Left: photo-$z$s. %The horizontal axis values are from the flexible model and the vertical axis values are from the frozen model. 
The black solid line represents the 1-to-1 relation of the two photo-$z$s, while the two black dashed lines show the 15 \% bias defined as $|\delta z|=|z_{\rm froz}-z_{\rm flex}|/(1+z_{\rm flex})$.
%, where we have used the apex froz and flex for frozen and flexible. 
Right: effective radius. 
%The axes and the solid line are defined as in the left panel. 
The solid line is again the 1-to-1 relation, while the dashed lines represent the 20\% bias defined as $|\delta r|=|r_{\rm froz}-r_{\rm flex}|/r_{\rm flex}$.}
%Right: The specific star formation rate (sSFR) versus mass for both the frozen model (blue) and flexible model (red). The solid orizontal line indicate the typical sSFR found for pBNs in CANDELS (\citealt{Huertas-Company2018ApJ...858..114H}) at redshift $1.4<z<2.0$, while the shaded regions indicate a $\pm0.4$dex errors, which are realistic statistical errors on sSFRs using 9-band KiDS photometry and photo-$z$ (see \citealt{Xie2023arXiv230704120X}). The numbers 1-6 represent the six nuggets in the order listed in Table \ref{tab:source_mod}.}
\label{fig:comparison}
\end{figure*}

In this appendix, we report the results of the ``frozen'' prior model described in \S\ref{modeling}, and compare them with the ``flexible'' prior results. In Table \ref{tab:source_mod_app}, we report the full multi-band parameters for the ``flexible'' model in Table \ref{tab:source_mod}. We can measure the mean and the standard deviation of the source magnitudes derived from the two models as a measure of the average consistency from the two sets of magnitudes, and find these to be $-0.002\pm 0.2139$. This shows no systematics between the two models and a statistical error consistent with typical photometric errors for ground-based observations. The corresponding photo-$z$s are also compared with the ones from the ``flexible'' model in the left panel in Fig. \ref{fig:comparison}: these also show a rather large scatter but within the typical uncertainties expected for ground-based photo-$z$s (e.g. \citealt{2020A&A...642A.200V, 2021inas.book..245A, 2022A&A...666A..85L}).
%To assess the impact of the variation on the photo-$z$s, we plot in the right panel of Fig. \ref{fig:comparison} the corresponding scatter on the source physical effective radii.
%However, we can have a sense of the impact of the new photo-$z$, looking at the corresponding scatter 
\ruili{
%In the right panel of Fig. \ref{fig:comparison} we plot the comparison of the physical effective radius (in kpc) and the corresponding error bars. We found that the effective radius from the two models agree with each other within 20\% bias. We note that this bias is mainly from the lensing model, rather than the uncertainties on photo-zs. We have determined that, for the worst case of photo-zs, J224546$-$295559, which has the largest photo-z bias (1.71-1.26=0.45) between the two models, this bias can contribute to only 1.6\% uncertainty.
In the right panel of Figure \ref{fig:comparison}, we compare the physical effective radii of the galaxies, measured in kpc, along with their respective error bars. Our analysis reveals that the effective radii derived from the two different models are in close agreement, with a discrepancy of $\lesssim$20\%. It is important to highlight that this deviation primarily comes from the lensing model itself, rather than the uncertainties associated with the photo-zs. To further illustrate this point, we examined the case of J224546-295559, which displays the most significant photo-z discrepancy between flexible model and frozen model,
%. Its difference in photo-z is 
$\Delta z=$0.45. Even in this extreme case, we have estimated that the impact of this photo-z discrepancy on the effective radius is minimal, contributing to only a 1.6\% uncertainty.}
%This is becuase, the typical uncertainties of the photo-zs is $\sim$ 0.2 %at z=1.5 and $\sim$ 0.45 at z=2. Assuming the worse case,  We have %calculated that the uncertainities on photo-zs would contraibute to %$\sim$1\% errors on the physical effective radius for a galaxy at $z=1.5$ %with photo-z uncertainties of $\sim 0.2$
%in the middle panel of the same 
%Finally, we cumulatively compare the stellar masses and the specific star formation rates in the right panel in Fig. \ref{fig:comparison}. 
\rui{Finally, we cumulatively compare the stellar masses and the sSFRs, which has been shown in \ref{fig:mass_ssfr} but has not discussed in deep.} As expected, the magnitude and photo-$z$ errors propagate more dramatically on the stellar population parameters, especially sSFRs, \rui{with a typical deviation of the order of $\sim 0.2$ dex for stellar masses and $\sim0.41$ dex for sSFRs. These are not dissimilar to typical statistical errors found for stellar population parameters using normal KiDS galaxies (e.g., $\sim 0.4$ dex, \citealt{Xie2023arXiv230704120X}),} although they are big enough to move galaxies above the upper limit assumed to separate compact star-forming systems from compact quenched systems in observations (i.e. $\rm \log (sSFR/Gyr)\sim0$, see again \citealt{Johnston2015MNRAS.453.2540J} and \citealt{Huertas-Company2018ApJ...858..114H}). 
%\rui{However, if we consider the typical errors of $\sim 0.4$ dex on sSFRs (marked as a shaded region) derived from 9-band KiDS photometry and photo-zs (see detail in \citealt{Xie2023arXiv230704120X}),}
\rui{However, all systems remain compatible with a 'quenching' scenario, as they are situated below the mean sequence of compact SFGs (see \ref{fig:mass_ssfr}).}
To be conservative, and considering the typical uncertainties on sSFRs (marked as grey shaded region) and the results of the “frozen” priors at face value, 2 of the 6 systems have both sizes and sSFRs that deviate from the expectation to be pBN (see a more detailed discussion in \S\ref{discussion}).
%To be conservative, taking the results of the ``frozen'' priors at face value, at least 2 of the 6 systems have both sizes and sSFRs that deviate from the expectation to be pBN (see a more detailed discussion in \S\ref{discussion}).  

\begin{table*}[h]
\begin{center}
\scriptsize
\caption{\label{tab:source_mod_app}
Source properties from the multi-band ray tracing ``frozen'' model.}
\begin{tabular}{l c c c c c c c c c c c c c c c c}
\hline \hline
ID & $R_{\rm eff}$ & $b/a$ & $pa$ & n & u & g & r & i & Z & Y & J & H & Ks \\
&(arcsec)  && (deg) & &(mag)&(mag)&(mag)&(mag)&(mag)&(mag)&(mag)&(mag)&(mag)\\
\hline
J023929$-$321129 & 0.051$\pm$0.002   & 0.48$\pm$0.03  & 90 & 2.6$\pm$0.1  & 23.84	& 23.46	&	23.30	&	23.34	&	23.29	&	23.46	&	23.17	&	22.73	&	-\\
J122456$+$005048* & 0.042$\pm$0.002  & 0.41$\pm$0.02  & 98   &2.0$\pm$0.1  &-&23.97	&	23.52	&	22.87	&	22.34	&	22.13	&	21.79	&	21.60	&	21.47 \\
J224546$-$295559&  0.056$\pm$0.005   & 0.45$\pm$0.06  & 16    &3.4$\pm$0.3  &24.10	&	23.81	&	23.71	&	23.75	&	23.70	&	23.15	&	22.57	&	-	&	-\\
J230226$-$335637  & 0.138$\pm$0.007   & 0.56$\pm$0.04  & 68   &2.4$\pm$0.3  &23.48	&	23.06	&	23.08	&	22.99	&	22.78	& 22.73	&	22.20	&	22.17	&	-\\
J232940$-$340922* & 0.077$\pm$0.012  &  0.29$\pm$0.03  & 168  & 1.8 $\pm$0.8  &-& 23.64	&	23.44	&	23.36	&	23.33	&	22.69	&	22.36	&	22.15	&	21.87\\
J234804$-$302855  & 0.167$\pm$0.008   & 0.38$\pm$ 0.03  & 71   &  1.3$\pm$0.2 &23.95	&	23.64	&	23.68	&	23.62	&	23.94	&	23.37	&	22.82	&	22.70	&	-\\
\hline \hline
\end{tabular}
\end{center}
\textsc{Note.} ---  Column content is as in Table \ref{tab:source_mod}.
%Column 1 is the KiDS ID, in the form of hh-mm-ss and deg-mm-ss. Columns 2-5 list the source parameters from the lensing model by assuming a S\'{e}rsic profile for the sources. From left to right are the $r-$band magnitude $r_{mag}$, the effective radius $R_{\rm eff}$ in the unit of arcsec, the minor-to-major axis ratio $b/a$, the position angle $pa$ and the S\'{e}rsic index $n$, respectively. Columns 6-14 list the magnitudes of each source obtained from lensing model by assuming a S\'{e}rsic profile for the sources. Errors in magnitudes are shown in Fig. \ref{fig:SED}.
\end{table*}

\end{appendix}



\end{document}